\documentclass[12pt]{article}
\usepackage{epsfig}\parskip 5pt plus 1pt
\usepackage{bbm}
\usepackage{amsmath}
\usepackage{amssymb}
\usepackage{amsfonts}
\usepackage{mathrsfs}
\usepackage{graphicx}
\newcommand{\raw}{\rightarrow}

\renewcommand{\>}{\rangle}
\newcommand{\be}{\begin{equation}}
\newcommand{\ee}{\end{equation}}
\newcommand{\bea}{\begin{eqnarray}}
\newcommand{\eea}{\end{eqnarray}}
\begin{document}
\begin{titlepage}
\vspace*{-1cm}
\flushright{FTUAM-07-4 \\ IFT-UAM/CSIC-07-10}
\vskip 2.0cm
\begin{center}
{\Large\bf
CP-violation from non-unitary leptonic mixing
}
\end{center}
\vskip 1.0  cm
\begin{center}
{\large E. Fern\'andez-Mart\'\i nez}$\,^a$~\footnote{enrique.martinez@uam.es},
{\large M.B. Gavela}$\,^a$~\footnote{belen.gavela@uam.es},
{\large J. L\'opez-Pav\'on}$\,^a$~\footnote{jacobo.lopez@uam.es}\\
\vskip .1cm
and {\large O. Yasuda}$\,^b$~\footnote{yasuda@phys.metro-u.ac.jp}
\\
\vskip .2cm
$^a\,$ Departamento de F\'\i sica Te\'orica and Instituto de F\'\i sica Te\'orica,
\\
Universidad Aut\'onoma de Madrid, 28049 Cantoblanco, Madrid, Spain
\vskip .1cm
$^b\,$ Department of Physics,\\
Tokyo Metropolitan University, Hachioji, Tokyo 192-0397, Japan
\end{center}
\vskip 0.5cm
\begin{abstract}
A low-energy non-unitary leptonic mixing matrix is a generic effect of a large class of theories accounting for neutrino masses. It is shown  how the extra CP-odd phases of a general non-unitary matrix
allow for sizeable CP-asymmetries in channels other than those dominant in the standard unitary case. The $\nu_\mu\rightarrow \nu_\tau$ channel turns out to be an excellent tool to further constrain moduli and phases. Furthermore, we clarify the relationship between our approach and the so-called
 ``non-standard neutrino interactions" schemes: the sensitivities explored here apply as well to such constructions.

\noindent
\end{abstract}
\end{titlepage}
\setcounter{footnote}{0}
\vskip2truecm
\newpage
\section{Introduction}
\label{intro}

  Non-zero neutrino masses are evidence for physics beyond the Standard Model (SM). The complete theory accounting for them and encompassing the Standard Model should be unitary, as mandated by probability conservation. The light fermionic fields in the theory may mix with other degrees of freedom. The complete mixing matrices are generically unitary. However, the effective $3 \times 3$ submatrices describing the mixing of the known fields need not to be unitary.

We will not particularize to any given model of new physics beyond the SM (BSM) and will assume that the total theory of the world is indeed unitary. Nevertheless, low-energy non-unitarity may result from BSM physics contributing to neutrino propagation, when the physical measurements are described solely in terms of SM fields.

 Low-energy non-unitarity may generically appear when the new fields of the BSM theory  are heavier than the electroweak scale $v$: this is the framework contemplated in this work. Consider
 a theory with scale $M\gg v$. The impact of those heavy fields at low energies is well described in terms of effective operators of dimension $5$ and higher, invariant under the SM gauge group, made out of the SM fields active at low energies and with coefficients weighted by inverse powers of the scale $M$.

 The only possible dimension $5$ operator of this type is the famous Weinberg operator~\cite{Weinberg:1979sa},
 \begin{equation}
 \delta{\cal L}^{d=5} = \frac{1}{2}\, c_{\alpha \beta}^{d=5} \,
\left( \overline{L^c}_{\alpha} \tilde \phi^* \right) \left(
\tilde \phi^\dagger \, L_{ \beta} \right) +  h.c.\,
\end{equation}
(where $c_{\alpha \beta}^{d=5}$ is the coefficient matrix of ${\cal{O}}(1/M)\,$),
which happens to violate $B-L$, an accidental symmetry of the SM. Upon electroweak symmetry breaking, $<\phi> =v$, this term results in Majorana neutrino masses. Such an operator is characteristic of all theories with Majorana neutrino masses, such as for instance the Seesaw model~\cite{seesaw}. It is very suggestive that the lowest-order effect of high-energy BSM physics may be neutrino masses, which indeed constitute  -together with dark matter- the first experimental evidence of particle physics BSM.

Such tiny neutrino masses may naturally result through the tree-level exchange of heavy particles, which may be either fermions or bosons. The exchange of heavy SM singlet fermions is the essence of  the minimal
Seesaw model (Type I Seesaw) and its generalizations. SM fermionic triplets may also mediate light neutrino masses~\cite{type-III}. Analogously, the exchange of heavy SM triplet scalars has been as well widely explored, as in the Type II Seesaw model and its generalizations~\cite{type-II}.
Theories with extra spatial dimensions may also induce very small  neutrino masses through tree-level exchanges of the Kaluza-Klein tower of singlet states~\cite{extradim}. Some supersymmetric frameworks have similar mechanisms. Furthermore,
 radiative models of neutrinos masses forbid tree-level exchanges while inducing Majorana masses at the loop level. Which ones among all these constructions result at low energies in non-unitary leptonic mixing?

 It is  expected in all generality that the tree-level exchange of heavy fermions (scalars) will (not) induce low-energy non-unitary contributions.
 In ref.~\cite{uni} it was argued that those theories inducing corrections to the low-energy lepton propagators are putative sources of a non-unitary leptonic mixing matrix. Specifically, whenever the  kinetic-energy terms for light leptons acquire flavor-dependent normalizations, wave function renormalization is required to recover canonically normalized kinetic energies, which often leaves as {\it collateral  damage} a non-unitary leptonic mixing matrix N, which replaces the usual unitary PMNS matrix, U$_{PMNS}$.

 Although the non-unitary mixing induced at low energies in the simplest Seesaw model is typically expected to be too small for detection in the near future, this is not necessarily true for variants of the Seesaw mechanism~\cite{Mohapatra:1986bd}, or other theories beyond the SM.
At the same time, we are about to enter an era of high precision in neutrino physics.  It is pertinent to ask whether such precision can shed further light on departures from unitarity,  which would probe the new physics behind.

  In ref.~\cite{uni} the so-called MUV (minimal unitarity violation) was developed and  the {\it absolute} values of the elements of the matrix N were determined, using data from neutrino oscillation experiments and weak decays. It turned out that non-unitary contributions, if present, are constrained to be smaller than the percent level. As for the {\it phases}, a non-unitary mixing $3\times 3$ matrix has  three phases in excess over those in the unitary case. No information on the size of the phases of the mixing matrix is available, neither on the standard ``unitary'' phases nor on the new non-unitary ones, as present oscillation data correspond mainly to disappearance experiments.

   It is the purpose of this work to explore the future sensitivity to the CP-odd phases of the leptonic mixing matrix, in case it turns out to be non-unitary. All oscillation channels will be analyzed. In particular, it will be shown that CP-asymmetries in the $\nu_\mu \rightarrow \nu_\tau$ channel are an excellent probe of such new physics. Notice that CP-odd effects in that channel are negligible in the standard unitary case, in which the golden channel for CP-violation is $\nu_e   \rightarrow  \nu_\mu$.  On the experimental side, our quest has led us to consider several future facilities. Because of the interest of tau detection, measurements at a Neutrino Factory~\cite{nf} will be considered in detail and favored over Super-Beams~\cite{sb} and $\beta$-Beams~\cite{bb}, even the highest energy ones.

   Some {\it a priori} different avenues for new physics explored in the literature  are the  ``non-standard neutrino interactions" or ``exotic neutrino interactions". These are usually implemented through the addition of effective four-fermion operators to the SM Lagrangian~\cite{newint,conchanir,yasuda}. These operators can affect the production and detection processes or modify the matter effects in the propagation. We will clarify the relationship between our framework and  those  proposals. The channels we will explore and the sensitivities we will predict will be shown to generically apply to them as well.

    Section~\ref{formalism} on Formalism attends to its name and also contains the main qualitative argument about the observability of effects related to non-unitarity. Sect.~\ref{sensitivity} explores the sensitivity to CP-odd effects induced in the $\nu_\mu \rightarrow \nu_\tau$
 and the  $\nu_e   \rightarrow  \nu_\mu$ channels. The comparison with the results in ``non-standard neutrino couplings"-scenarios is performed in Sect.~\ref{non-standard}. In Sect.~\ref{conclusions} we conclude. Finally, apendix A introduces a formalism to derive oscillation probabilities in matter with constant density, in the MUV scheme.

\section{Formalism}
\label{formalism}
Let us parameterize the general non-unitary matrix $N$, which relates flavor and mass fields\footnote{Throughout the paper, Greek (Latin) indices label the flavor (mass) basis.}
\be
\label{field-transf1}
\nu_{\alpha} = N_{\alpha i}\, \nu_{i}  \, ,
\ee
 as the product  of an hermitian and a unitary matrix, defined by
\be
N\equiv(1+\eta) U ,
\label{N}
\ee
\noindent
with $\eta^\dagger=\eta$. The bounds derived in Ref.~\cite{uni} for the modulus of the elements of $NN^\dagger$ from universality tests, rare lepton decays and the
invisible width of the $Z$, also apply to the elements of $\eta$, since
$NN^\dagger=(1+\eta)^2\approx 1+2\eta$ and it follows that
\medskip
\begin{eqnarray}
|\eta| =
\begin{pmatrix}
 |\eta_{ee}|<5.5\cdot 10^{-3}  & |\eta_{e\mu}| < 3.5 \cdot 10^{-5}  &    |
\eta_{e\tau}|< 8.0 \cdot 10^{-3} \\
 |\eta_{\mu e}|< 3.5 \cdot 10^{-5}  & |\eta_{\mu\mu}|<5.0\cdot 10^{-3}  &    |
\eta_{\mu\tau}| < 5.1 \cdot 10^{-3} \\
 |\eta_{\tau e}| < 8.0 \cdot 10^{-3} & |\eta_{\tau\mu}| < 5.1 \cdot 10^{-3}
&  |\eta_{\tau\tau}|<5.0\cdot 10^{-3}
\end{pmatrix} \,,
\label{limits}
\end{eqnarray}
at the $90\%$ confidence level. The bound on $\eta_{\mu \tau}$ has been updated with the latest experimental bound on $\tau \raw \mu\gamma$~\cite{taumugamma}. Eq.~(\ref{limits}) shows that the matrix $N$ is constrained to be unitary, within a $10^{-2}$ accuracy or better. The unitary matrix $U$ in Eq.~(\ref{N}) can thus be identified with the usual unitary mixing matrix $U=U_{PMNS}$, within the same accuracy. The flavor
eigenstates can then be conveniently expressed as\footnote{As neutrino masses are forbidden in the SM, the handy superscript $SM$ is an abuse of language, that we allow ourselves to describe the flavor eigenstates of the standard unitary analyzes.}

\be
|\nu_\alpha> = \dfrac{(1+\eta^*)_{\alpha \beta}U^*_{\beta i}}{\left[ 1+2\eta_{\alpha\alpha}+(\eta^2)_{\alpha\alpha}\right]^{1/2}}\,|\nu_i> \equiv \dfrac{(1+\eta^*)_{\alpha \beta}}{\left[ 1+2\eta_{\alpha\alpha}+(\eta^2)_{\alpha\alpha}\right]^{1/2} }\,|\nu^{SM}_\beta> .
\label{estados}
\ee
It follows that the neutrino oscillation amplitude, neglecting terms quadratic in $\eta$, is given simply by
\be
<\nu_\beta|\nu_\alpha (L)> =  A^{SM}_{\alpha \beta}(L)\,\left(1-\eta_{\alpha\alpha}-\eta_{\beta\beta} \right) +\sum_\gamma \left( \eta^*_{\alpha \gamma}A^{SM}_{\gamma \beta}(L)+
\eta_{\beta \gamma}A^{SM}_{\alpha \gamma}(L) \right)\,,
\label{amplitud}
\ee
\noindent
with
\be
A^{SM}_{\alpha \beta}(L)\equiv<\nu^{SM}_\beta|\nu^{SM}_\alpha (L)>
\ee
\noindent
being the usual oscillation amplitude of the unitary analysis.

New CP-violation signals arising from the new phases in $\eta$, require to contemplate appearance channels, $\alpha\ne\beta$.  The best sensitivities to such phases will be achieved in a regime where the first term in Eq.~(\ref{amplitud}) is suppressed. This happens at short enough baselines, where the standard appearance amplitudes become vanishingly small while $A^{SM}_{\alpha \alpha}(L)\simeq 1$. The total amplitude is then well approached by
\be
<\nu_\beta|\nu_\alpha (L)> =  A^{SM}_{\alpha \beta}(L)+ 2\eta_{\alpha \beta}^* + \mathcal{O}(\eta\,A),
\label{amp}
\ee
where $\mathcal{O}(\eta\,A)$ only includes appearance amplitudes and $\eta\,$ components with flavor indices other than $\alpha \beta$.
This is an interesting property as it implies that, at short enough baselines, each oscillation probability in a given flavor channel, $P_{\alpha \beta}$, is most sensitive to the corresponding
$\eta_{\alpha \beta}$. The other elements of the $\eta$ matrix can be safely disregarded in the analyzes below, without implying to assume zero values for them. That is, their effect is generically subdominant, a fact that will be numerically checked for the main contributions, as explained later on. This also means that the subleading corrections from the cross sections and fluxes discussed in Ref.~\cite{uni}, which induce also $\mathcal{O}(\eta\,A)$ corrections, do not need to be taken into account.

For instance, in a two family scenario and within the above-described approximation, the oscillation probability would read:
\be
P_{\alpha \beta} = \sin^{2}(2\theta)\sin^2\left(\frac{\Delta L}{2}\right)
-4|\eta_{\alpha \beta}|\sin\delta_{\alpha \beta}\sin(2\theta)\sin\left(\frac{\Delta L}{2}\right)
+4|\eta_{\alpha \beta}|^2,
\label{prob}
\ee
\noindent
where $\Delta = \Delta m^2 /2E$ and $\eta_{\alpha \beta}=|\eta_{\alpha \beta}|e^{-i\delta_{\alpha \beta}}$.
The first term in Eq.~(\ref{prob}) is the usual oscillation probability when the mixing matrix is unitary. The third term is
the zero-distance effect stemming from the non-orthogonality of the flavor eigenstates. Finally, the second term is the CP-violating interference between the other two. Notice that the latter is linearly sensitive to both phases and moduli, a fact which will be at the origin of the improvement in the sensitivity to the moduli, for non-trivial values of the phases $\delta_{\alpha \beta}$, as compared to previous analyzes in the literature. In this same two-family scenario, a CP-asymmetry can be written as
\begin{equation}
\mathcal{A}^{CP}_{\alpha\beta}=\frac{P_{\alpha\beta}-P_{\bar\alpha \bar\beta}}{P_{\alpha\beta}+P_{\bar\alpha \bar\beta}}\sim \frac{ -4|\eta_{\alpha \beta}|\sin\delta_{\alpha \beta}}{\sin(2\theta)\sin\left(\frac{\Delta L}{2}\right)}\,,
\label{asym}
\end{equation}
where, for illustration, it is implicitly assumed that we work in a regime in which the term quadratic in $\eta$ in Eq.~(\ref{prob}) is negligible with respect to the first term, that is, with respect to the standard contributions.

In the sections below, we will analyze the new sources of CP violation in the $\nu_e \rightarrow \nu_\tau$ and the $\nu_\mu \rightarrow \nu_\tau$ appearance channels, since present constraints on $\eta_{e \mu}$ are too strong to allow a signal in the $\nu_e \rightarrow \nu_\mu$ channel (see Eq.~(\ref{limits})).  When numerically computing a given $P_{\alpha \beta}$, the only approximation performed will be to neglect all $\eta$ elements but that corresponding to the channel under consideration, $\eta_{\alpha \beta}$. They should be indeed subdominant, as
illustrated by Eq.~(\ref{amp}).
Furthermore, we have checked this approximation as follows. The numerical fits have been performed in two ways. First, setting to zero all the elements of $\eta$ except $\eta_{\alpha\beta}$. Next, allowing the other off-diagonal elements of $\eta$ to vary (except for $\eta_{e \mu}$, which is extremely well constrained).
The results are indistinguishable within the accuracy explored.
The effect of the diagonal elements is expected to be even smaller since they cannot induce CP asymmetries.

Finally, the parameterization of $N$ in Eq.~(\ref{N}) is on purpose very similar to that used to
study non-standard neutrino interactions from four-fermion operators in Refs.~\cite{newint,conchanir,yasuda}. Some CP-odd effects have also been considered in these scenarios~\cite{conchanir}.  We leave to Section~\ref{non-standard}  the task of clarifying the similarities and differences with our framework, their equivalence in certain regimes and the range of application of our numerical bounds to those constructions.

\section{Sensitivity to the new CP-odd phases}
\label{sensitivity}
As suggested by Eqs.~(\ref{prob}) and (\ref{asym}), the best sensitivities to  CP-violation  will be achieved at
short baselines and high energies, where the standard term is suppressed by $\sin^2(\frac{\Delta L}{2})$. We will therefore
study a Neutrino Factory beam resulting from the decay of $50$ GeV muons, to be detected at a $130$ Km baseline, which matches for
example the CERN-Frejus distance.  For these values, $\sin(\frac{\Delta_{31} L}{2})\simeq 1.7\cdot10^{-2}$ and $\sin(\frac{\Delta_{21} L}{2})\simeq 6\cdot10^{-4}$, where $\Delta_{jk}\equiv(m^2_j-m^2_k)/2E$.
All terms in the  oscillation probability Eq.~(\ref{prob}) can then be of similar order for the channels $\nu_\mu \raw \nu_\tau$ and $\nu_e \raw \nu_\tau$, if the corresponding $\eta_{\alpha\beta}$ values are close to their experimental
limits in Eq.~(\ref{limits}).
  In what follows, we will assume $2\cdot10^{20}$ useful decays per year and five
years running with each polarity.

The appearance of $\nu_\tau\,$s will be contemplated assuming a 5 Kt Opera-like detector.
The efficiencies and backgrounds for the measurement of $\nu_e\rightarrow\nu_\tau$ transitions at a Neutrino
Factory have been taken  from Ref.~\cite{silver}.
 A similarly detailed analysis for the $\nu_\mu \raw \nu_\tau$ channel is still lacking in the literature. In fact, $\nu_\mu \raw \nu_\tau$ oscillations are one of the main backgrounds in the detection of
$\nu_e\rightarrow\nu_\tau$ transitions.
 As the signal for the former transition is stronger, looser cuts should be sufficient when analyzing it. Namely, since charge identification should not be mandatory, it may be possible to
study not only the $\tau$ into $\mu$ decay mode, but also the rest of the $\tau$ decay modes, gaining a factor of 5 in sensitivity which, on the other side, will mean larger backgrounds. We have thus considered sensitivities and
backgrounds a factor 5 larger when analyzing the $\nu_\mu \raw \nu_\tau$  channel
than those used for the $\nu_e\rightarrow\nu_\tau$ channel~\cite{migliozzi}.

In the numerical analysis, the complete oscillation probabilities in matter have been
used (albeit with the simplifications on $\eta$ previously described). It is interesting, though, to understand qualitatively and in detail  the role of matter effects and, whenever relevant, the dependence on
 small neutrino parameters, such as $\Delta_{21}L$ and $\sin2\theta_{13}$. For this purpose, we will consider below an expansion of the oscillation probabilities
   at higher order than that implied by Eq.~(\ref{prob}). Taking into account that $\Delta_{31}L\sim AL\sim10^{-2}$, where $A=\sqrt{2}G_F n_e$, with $G_F$ being the Fermi constant and $n_e$ the electron density in the Earth crust, $\Delta_{21}L\sim10^{-3.5} $, $\sin2\theta_{13}\lesssim10^{-0.5}$, $|\eta_{\alpha\beta}|\lesssim10^{-2}$ and $|\eta_{e\mu}|<3.6\cdot10^{-5}$, it is consistent to expand to second order in the following parameters:
   \be
   \sin^22\theta_{13}\,, \,\,\Delta_{21}L\,,\,\,(\Delta_{31}L)^2\,, \,\,(AL)^2\,,\,\,\eta_{\alpha\beta},
 \label{parameters}
 \ee
 with $\eta_{e\mu}$ set to zero, as its contributions are always suppressed by extra small parameters.

\subsection{The $\nu_\mu \raw \nu_\tau$ channel}

The expression for $P_{\mu\tau} $ expanded to the order just described can be found in \ref{appendix1}. Matter effects are subleading, as this probability is not suppressed by small standard parameters such as
$\sin\theta_{13}$ or $\Delta_{12}$. The two family approximation in Eq.~(\ref{prob}) is thus  very accurate to understand qualitatively the results and reads
\be
P_{\mu \tau} = \sin^{2}(2\theta_{23})\sin^2\left(\frac{\Delta_{31} L}{2}\right)
-2|\eta_{\mu \tau}|\sin\delta_{\mu \tau}\sin(2\theta_{23})\sin\left(\Delta_{31} L\right)
+4|\eta_{\mu \tau}|^2.
\label{probmutau}
\ee
\noindent
This equation indicates that the CP-odd interference term is only suppressed linearly in $|\eta_{\mu \tau}|$.
\begin{figure}[t]
\vspace{-0.5cm}
\begin{center}
\begin{tabular}{cc}
\hspace{-0.55cm} \includegraphics[width=7.5cm]{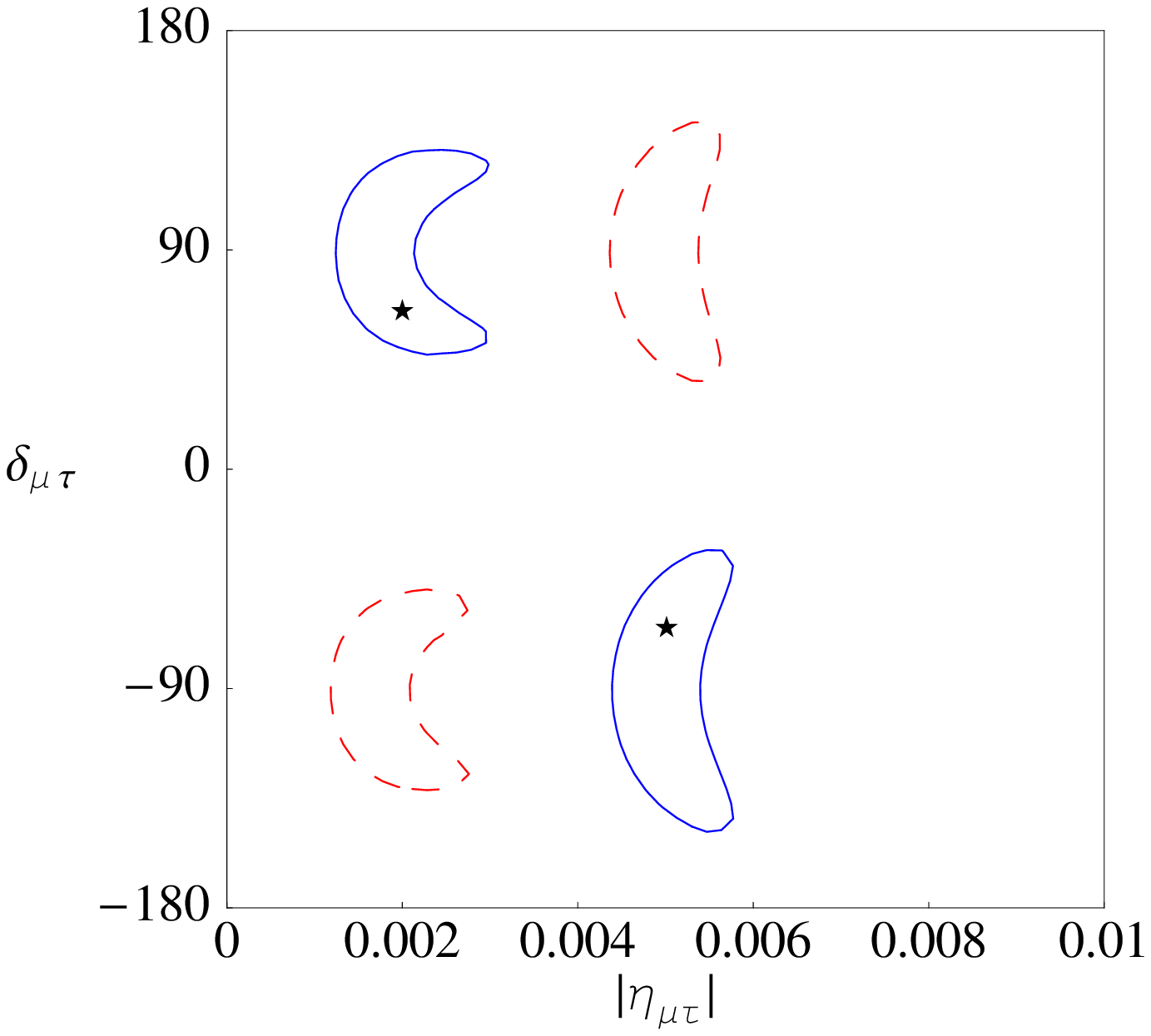} &
		 \includegraphics[width=7.5cm]{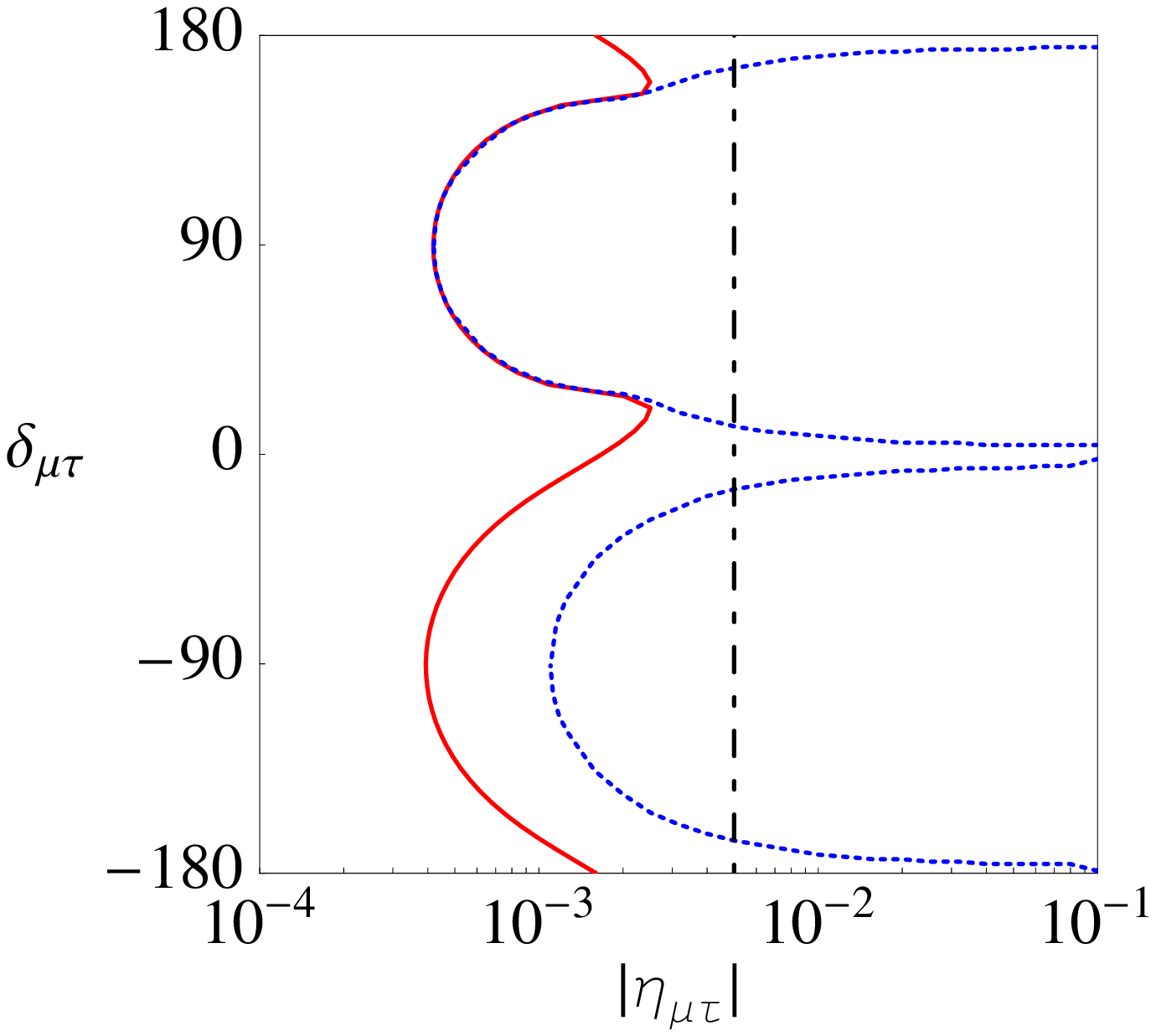}
\end{tabular}
\caption{\it Left: $3 \sigma$ contours for two input values of $|\eta_{\mu \tau}|$ and $\delta_{\mu \tau}$ represented by
the stars. Right: the
solid line represents the $3\sigma$ sensitivity to $|\eta_{\mu \tau}|$ as a function of $\delta_{\mu \tau}$, the dotted
line the $3\sigma$ sensitivity to $\delta_{\mu \tau}$ and the dotted-dashed line represents
the present bound from $\tau \raw \mu \gamma$.}
\label{fig:mutau}
\end{center}
\end{figure}
This can indeed be observed in the result of the complete numerical computation, Fig.~\ref{fig:mutau}, which shows the sensitivities to $|\eta_{\mu \tau}|$ and $\delta_{\mu \tau}$ obtained. The left
panel represents two fits to two different input values of $|\eta_{\mu \tau}|$ and $\delta_{\mu \tau}$ (depicted by stars). The dashed lines correspond to fits done assuming the {\it wrong} hierarchy, that is the opposite sign for $\Delta_{31}$ to that with which the number of events were generated. As expected from Eq.~(\ref{probmutau}), a change of sign for the mass difference can be traded by a change of sign
for $\delta_{\mu \tau}$. Nevertheless, this does not spoil the potential for the discovery of CP violation, since a non-trivial value for  $|\delta_{\mu \tau}|$ is enough to indicate CP violation.
 Furthermore, the sinusoidal dependence implies as well a degeneracy between $\delta_{\mu \tau} \raw 180^\circ-\delta_{\mu \tau}$, as reflected  in  the figure.

The right panel in Fig.~\ref{fig:mutau} depicts the $3\sigma$ sensitivities to $|\eta_{\mu \tau}|$ (solid line) and $\delta_{\mu \tau}$ (dotted line), while the present bound from $\tau \raw \mu \gamma$ is also shown (dashed line). The poorest sensitivity to $|\eta_{\mu \tau}|$,
around $10^{-3}$, is found in the vicinity of $\delta_{\mu \tau} = 0$ and $\delta_{\mu \tau} = 180^\circ$, where the CP-odd interference term
 vanishes and the bound is placed through the subleading $|\eta_{\mu \tau}|^2$ term. The latter is also present at zero distance and its effects were already considered in Ref.~\cite{uni}, obtaining a bound of similar magnitude.
 The sensitivity to $|\eta_{\mu \tau}|$ peaks around $|\eta_{\mu \tau}| \simeq 4\cdot 10^{-4}$ for
$\delta_{\mu \tau} \simeq \pm 90^\circ$, where $\sin\delta_{\mu \tau}$ is maximum. That is, for non-trivial values of $\delta_{\mu \tau}$
not only CP-violation could be discovered, but values of $|\eta_{\mu \tau}|$ an order of magnitude smaller could be probed.

\subsection{The $\nu_e \raw \nu_\tau$ channel}

Contrary to $P_{\mu \tau}$, all terms in $P_{e \tau}$, standard and new ones, are suppressed by at least one of the two small standard parameters $\sin\theta_{13}$ and $\Delta_{12}$.  The oscillation probability, expanded to second order in
 the parameters in Eq.~(\ref{parameters}), reads

\bea
P_{e\tau}&=&c^2_{23}\sin^22\theta_{13}\left(\frac{\Delta_{31}L}{2}\right)^2+s^2_{23}\sin^22\theta_{12}\left(\frac{\Delta_{21}L}{2}\right)^2\nonumber
\\
&-&c_{13}\sin2\theta_{12}\sin2\theta_{23}\sin2\theta_{13}\cos\delta
\left(\frac{\Delta_{21}L}{2}\right)\left(\frac{\Delta_{31}L}{2}\right) \nonumber\\
&-&2|\eta_{e \tau}|c_{23}\sin2\theta_{13}\sin(\delta+\delta_{e \tau})(\Delta_{31}L)
\nonumber\\
&+&2|\eta_{e \tau}|s_{23}c_{13}\sin2\theta_{12}\sin\delta_{e \tau}\sin(\Delta_{21}L)\nonumber\\
&+&4|\eta_{e \tau}|^2\,,
\label{probetau}
\eea
where $s_{ij}$, $c_{ij}$ stand for $\cos\theta_{ij}$, $\sin\theta_{ij}$, respectively.

The first three terms in Eq.~(\ref{probetau}) are the usual unitary contributions, suppressed quadratically in $\sin\theta_{13}$ and $\Delta_{12}$.
 The next two are interference terms between
the unitary oscillation contribution and the non-unitarity parameter $\eta_{e \tau}$, and are CP-odd and suppressed~\footnote{ This dependence was previously observed for $\nu_e\rightarrow\nu_\mu$ transitions in a related context in Ref.~\cite{conchanir}.} by $|\eta_{e \tau}|$ and
either $\sin\theta_{13}$ or $\Delta_{12}$.
 The last term is the zero distance effect only proportional to $|\eta_{e \tau}|^2$. Notice that Eq.~(\ref{probetau})
would also be valid for the $P_{e \mu}$ oscillation probability replacing $s_{23} \raw -c_{23}$, $c_{23} \raw s_{23}$ and
$\eta_{e \tau} \raw \eta_{e \mu}$.
Although in the numerical analysis we have used the full oscillation probability in matter, the approximation in which
Eq.~(\ref{probetau}) has been obtained shows no sensitivity to matter effects.

\begin{figure}[t]
\vspace{-0.5cm}
\begin{center}
\begin{tabular}{cc}
\hspace{-0.55cm} \includegraphics[width=7.5cm]{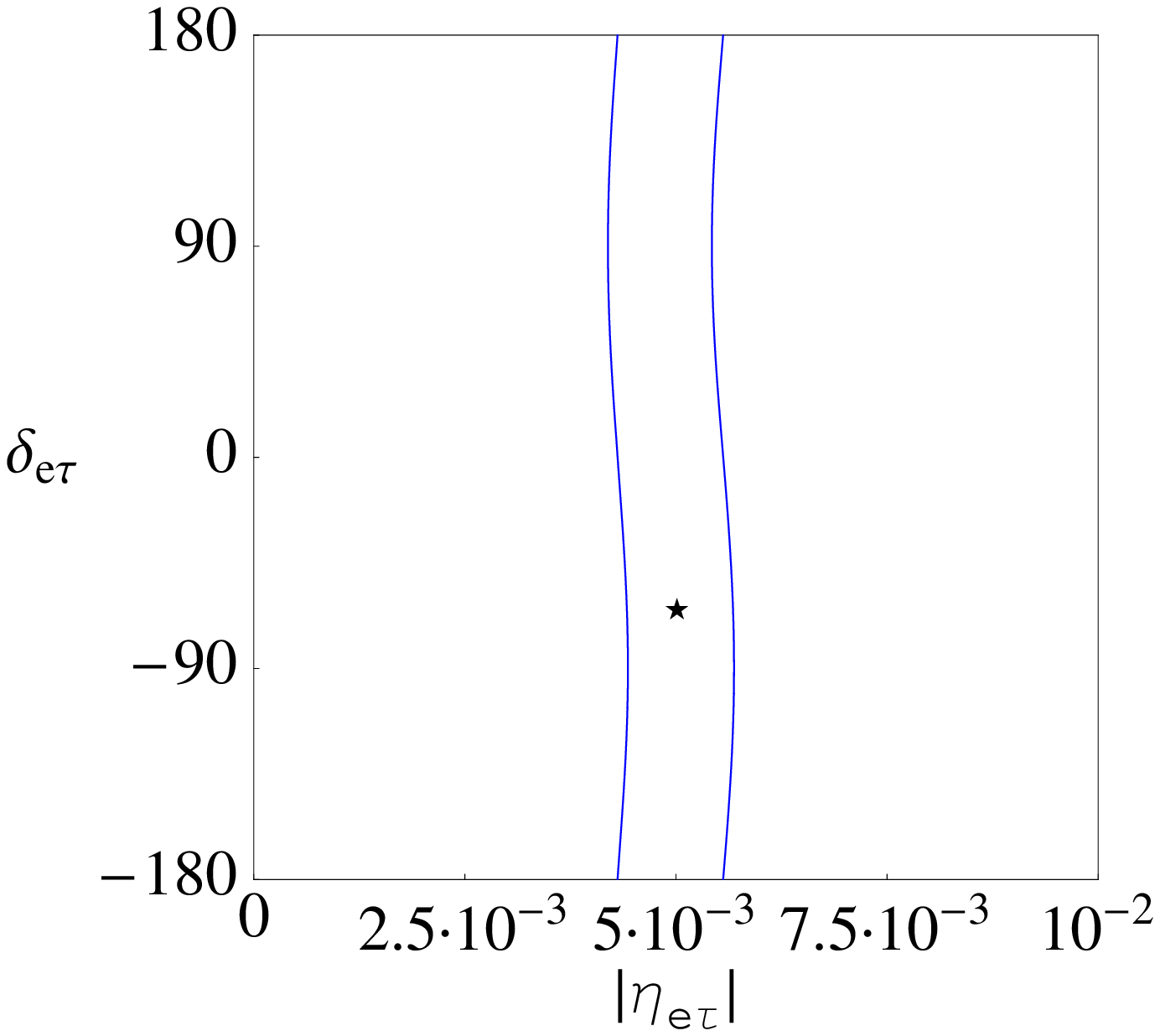} &
		 \includegraphics[width=7.5cm]{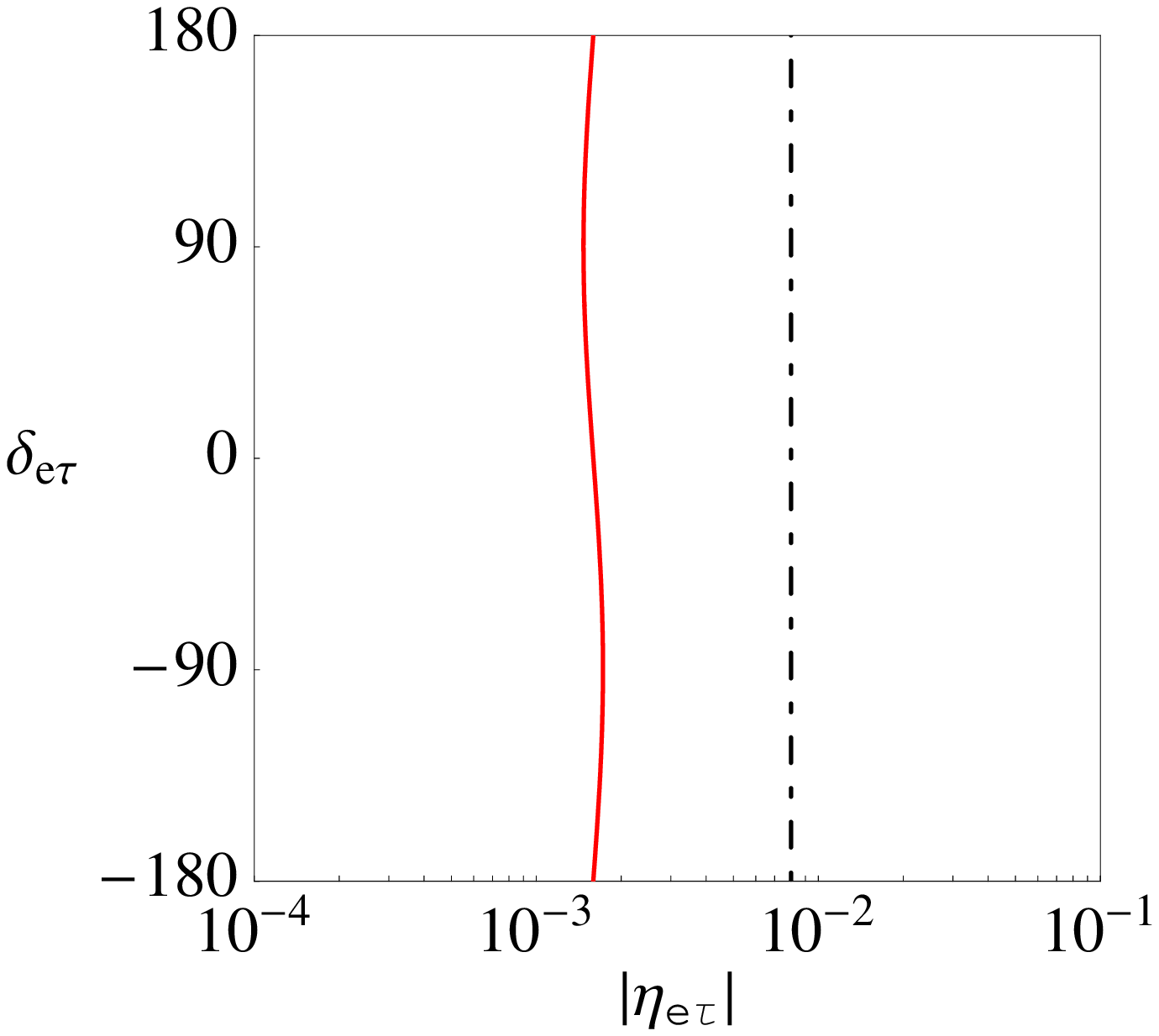}
\end{tabular}
\caption{\it Left: $3 \sigma$ contour for an input value of $|\eta_{e \tau}|$ and $\delta_{e \tau}$ represented by the star; the $|\eta_{e \tau}|^2$ term dominates and there is no sensitivity to the CP-violating phase. Right: $3 \sigma$ sensitivity to $|\eta_{e \tau}|$ as a function of $\delta_{e \tau}$, the dashed line represents the present bound from $\tau \raw e \gamma$.}
\label{fig:etau}
\end{center}
\end{figure}
Fig.~\ref{fig:etau} shows the sensitivities to $|\eta_{e \tau}|$ and $\delta_{e \tau}$ with this setup.
The left panel depicts a fit to the input value represented by the star: the dependence on $\delta_{e \tau}$ is seen to be very mild and no measurement of this quantity can be performed. This is easily understood from Eq.~(\ref{probetau}), for $|\eta_{e \tau}|>10^{-3}$: the last term, proportional to
$|\eta_{e \tau}|^2$, dominates over the CP-violating ones and no information on $\delta_{e \tau}$ can be extracted.
In consequence, in the right
panel we find $3 \sigma$ sensitivities to $|\eta_{e \tau}|$ around $10^{-3}$, but no sensitivity to $\delta_{e \tau}$.
The sensitivity to $|\eta_{e \tau}|$ from the $|\eta_{e \tau}|^2$ term is 
also present at zero distance and already studied in Ref.~\cite{uni}, where a similar bound was obtained. We have checked that increasing the statistics
by a factor 100, values of $|\eta_{e \tau}|$ an order of magnitude smaller would be accessible and the CP-violating terms would start
to dominate, providing sensitivity to $\delta_{e \tau}$.

\section{Non-unitarity vs non-standard interactions}
\label{non-standard}
Non-standard neutrino interactions are usually introduced through the addition of effective four-fermion operators to the SM Lagrangian~\cite{newint,conchanir,yasuda}, made out of the SM fields and invariant under the SM gauge group. They can affect the production or  detection processes or modify the matter effects in the propagation, depending on the operator or combination of operators considered.

In order to clarify the impact of each operator, let us first  write the effective SM Lagrangian after integrating out the $W$ field~\footnote{For simplicity, we do not show the neutral current interactions nor quark sector, since they are not relevant for this discussion.}. In the flavor basis, it reads
\be
-{\cal L}^{CC}=\frac{4G_F}{\sqrt{2}}[\sum_{\alpha\beta}(\bar\nu_{\alpha}^{SM} \gamma_{\mu} P_L l_{\alpha})
(\bar d_{\beta} \gamma_{\mu} P_L u_{\beta})\,+
\,
(\bar\nu_{\alpha}^{SM} \gamma_{\mu} P_L l_{\alpha})
(\bar l_{\beta} \gamma_{\mu} P_L \nu_{\beta}^{SM})\,+\,h.c.]\,.\,
\label{LCCSM}
\ee
The first term in Eq.~(\ref{LCCSM}) contributes to hadronic detection and production processes such as $\nu_\alpha d\rightarrow l_\alpha^{-} u $. The second term describes leptonic production and detection processes involving
two neutrinos, such as muon decay at the NF, $\mu^{+}\rightarrow e^{+}\nu_e\bar\nu_\mu $, as well as matter effects for $\alpha = \beta=e$.

Non-standard neutrino interactions are usually introduced through analogous effective operators that can be non-diagonal in flavor. For instance, in Ref.~\cite{yasuda} new matter effects stemming from the operator $\sim\left( \nu_{\alpha}^{SM} \gamma_{\mu} P_L\nu_{\beta}^{SM}\right)\left(\bar f\gamma_{\mu}f \right)$ were studied. This kind of operators give rise to non-diagonal corrections to the effective potential for the evolution equation in matter (in the SM flavor basis).

In Ref.~\cite{conchanir}, another kind of four-fermion operators were introduced:
\bea
{ O}^{P}&=&\frac{-4}{\sqrt{2}}\sum_{\alpha\neq e}G_{e\alpha}^{P}\left( \bar\mu \gamma_{\mu} P_L \nu_{\mu}^{SM}\right) \left( \bar\nu_{\alpha}^{SM} \gamma_{\mu} P_L e\right)\,+\,h.c.\,,\nonumber
\\
{ O}^{D}&=&\frac{-4}{\sqrt{2}}\sum_{\beta\neq \mu}G_{\mu\beta}^{D}\left( \bar\nu_{\beta}^{SM} \gamma_{\mu} P_L \mu\right)\left( \bar d \gamma_{\mu} P_L u\right)\,+\,h.c.\,,
\label{OPConcha}
\eea
\noindent
which affect the production ($O^{P}$) and detection ($O^{D}$) processes at a Neutrino Factory, but do not correct matter effects. In this case, the relevant terms in the interaction Lagrangian are modified as follows:
\bea
-{\cal L}^{int}= &&\frac{4G_F}{\sqrt{2}}\sum_{\alpha}\biggr[ \left( \delta_{\alpha\mu}+\frac{G_{\mu\alpha}^{D}}{G_F}\right) \left( \bar\nu_{\alpha}^{SM} \gamma_{\mu} P_L \mu\right) \left( \bar d \gamma_{\mu} P_L u\right)\,+\,h.c. \nonumber
\\ &+&
\left( \delta_{\alpha e}+\frac{G_{e\alpha}^{P}}{G_F}\right)\left( \bar\nu_{\alpha}^{SM} \gamma_{\mu} P_Le\right) \left( \bar\mu \gamma_{\mu} P_L \nu_{\mu}^{SM}\right)+h.c.\biggl] .
\label{Lconcha}
\eea
Defining now $\nu_\alpha^{SM} \equiv U_{\alpha i}\,\nu_i$, with U being the PMNS matrix, the effective production and detection states are given by:
\bea
|\nu_e^{p}> &=& (1+\epsilon^{*p})_{e\beta}|\nu^{SM}_\beta>=(1+\epsilon^{*p})_{e\beta}U_{\beta i}^* |\nu_i> \nonumber
\\
|\nu_\mu^d> &=& (1+\epsilon^{*d})_{\mu\beta}|\nu^{SM}_\beta>=(1+\epsilon^{*d})_{\mu\beta}U_{\beta i}^* |\nu_i>,
\label{estadosconcha}
\eea
where $\epsilon^{p(d)}_{\alpha\beta}=\dfrac{G_{\alpha\beta}^{P(D)}}{G_F}$, up to normalization factors.

These expressions are very similar to our parameterization of the effects of a non-unitary mixing matrix, Eq.~(\ref{estados}). In fact, the latter can also be encoded in terms of effective four-fermion operators, after integrating out the $W$ and $Z$ bosons. The difference is that, in the case of a non-unitary mixing matrix, the coefficients of the different effective operators induced by it and contributing to production, detection and matter effects are not independent but related.
For instance, it follows from Eq.~(\ref{N}) that $\eta_{\alpha\beta}=\eta_{\beta\alpha}^*$. Which would mean $\epsilon_{\alpha\beta}^p = \epsilon_{\beta \alpha}^{d*}$, a constraint usually not required when introducing non-standard interactions. When such equality holds,
 the oscillation physics induced by non-standard interactions is equivalent in vacuum to that stemming from  non-unitarity in the MUV scheme. 
Furthermore, even if no such relations among  $\epsilon_{\alpha\beta}^p$ and $\epsilon_{\alpha \beta}^d$ are assumed, the order of magnitude of the bounds obtained in the previous chapters should apply as well to non-standard interactions, barring very fine tuned cancellations. The new CP signals analyzed in this work are also probes of the phases of non-standard interactions.

Finally, in Ref.~\cite{uni} it was shown that matter effects are also modified in a very definite way in the presence of a non-unitarity mixing matrix. These new matter effects are not independent from the production/detection effects, contrary to the customary assumption in studies of  non-standard neutrino interactions. Nevertheless, as we have studied a setup such that matter effects - standard and new ones - are negligible, the bounds derived here can be applied directly to the case of non-standard interactions.

\section{Conclusions}
\label{conclusions}

     It is  important to analyze low-energy data without assuming a unitary leptonic mixing matrix, as it is a generic window on new physics, even if the effects are expected to be extremely tiny in the simplest models of neutrino masses.

  A non-unitary matrix has not only  more moduli than a unitary one, but also supplementary phases which may lead to new signals of CP-violation. In particular, an asymmetry between the strength of $\nu_\mu\rightarrow \nu_\tau$ oscillations  versus that for $\bar\nu_\mu\rightarrow \bar \nu_\tau$ has been shown to be a beautiful  and  excellent probe of new physics, when measured at short-baselines ($\sim 100$ km.) using a Neutrino Factory  beam of energy $\mathcal{O}(20 GeV)$. Non-trivial values of the new phases can  lead not only to the discovery of
  CP-violation associated to the new physics, but also allow to probe the absolute value of the moduli down to $10^{-4}$. This means an improvement of an order of magnitude over previous analyzes of future facilities, which were sensitive only to the moduli.

  Our analyzes for future sensitivities to non-unitarity, as well as the new signals of CP-violation explored here, have been shown to also apply to the physics of ``non-standard neutrino interactions", barring extremely fine-tuned cancellations.

\appendix
\def\thesection{Appendix \Alph{section}}
\section{$P_{\mu\tau}$ expansion\label{appendix1}}

In this Appendix, a convenient formalism is introduced
to derive oscillation probabilities $P_{\alpha \beta}$ in matter with constant density,
in the MUV scheme, and the results are then applied to the $\mu \raw \tau$ channel.

Several years ago Kimura, Takamura and Yokomakura derived a nice compact
formula~\cite{Kimura:2002hb}
for the neutrino oscillation probability in matter with
constant density. They showed that
the quantity $\tilde{U}^\ast_{\alpha j}\tilde{U}_{\beta j}$
in matter
can be expressed as a linear combination of the quantity
$U^\ast_{\alpha j}U_{\beta j}$ in vacuum, where $U_{\alpha j}$ and
$\tilde{U}_{\alpha j}$ stand for the matrix element of the
PMNS matrix in vacuum and in matter, respectively.
A simple derivation of their formula and its generalization are given
in Ref.~\cite{yasuda-kty}.  Here it is shown that the framework in
Ref.~\cite{yasuda-kty} can be applied also to the MUV case.

Time evolution of the neutrino mass eigenstate
$\Psi_m\equiv(|\nu_1\> ,|\nu_2\>,|\nu_3\> )^T$ in matter with constant
density in the MUV framework is given by
\begin{eqnarray}
i{\frac{d\Psi_m(t)}{dt}}=
({\cal E}
+N^T{\cal A}N^{*}
)\,\Psi_m(t),
\label{nonunitary1}
\end{eqnarray}
where
${\cal E}\equiv{\mbox{\rm diag}}(E_1,E_2,E_3)$
is the energy matrix in the mass eigenstate in vacuum,
${\cal A}\equiv\sqrt{2}G_F{\mbox{\rm diag}}
(n_e-n_n/2,-n_n/2,-n_n/2)$,
$n_n$ is the neutron density,
and the non-unitary matrix $N$ relates the flavor fields to
the mass fields (cf. Eq.~(\ref{field-transf1})). Diagonalizing the hermitian matrix ${\cal E}+N^T{\cal A}N^{*}$
with a unitary matrix $W$
\begin{eqnarray}
{\cal E}+N^T{\cal A}N^{*}=W{\cal \tilde{E}}W^{-1},
\label{energymatrix}
\end{eqnarray}
where ${\cal \tilde{E}}\equiv {\mbox{\rm diag}}
(\tilde{E}_1,\tilde{E}_2,\tilde{E}_3)$
is the energy matrix in matter
in the MUV scheme, the mass eigenstate at distance $L$ can be solved as
\begin{eqnarray}
\Psi_m(L)=W\exp(-i{\cal \tilde{E}}L)W^{-1}\Psi_m(0).
\label{psim}
\end{eqnarray}
We define the modified amplitude
$\hat{A}(\nu_\alpha\rightarrow\nu_\beta)
\equiv A(\nu_\alpha\rightarrow\nu_\beta)
(NN^\dagger)_{\alpha\alpha}^{1/2}
(NN^\dagger)_{\beta\beta}^{1/2}$, 
\begin{eqnarray}
\hat{A}(\nu_\alpha\rightarrow\nu_\beta)
=[N^{*}W\exp(-i{\cal \tilde{E}}L)W^{-1}N^T]_{\alpha\beta},
\nonumber
\end{eqnarray}
and the modified probability
$\hat{P}(\nu_\alpha\rightarrow\nu_\beta)
\equiv|\hat{A}(\nu_\alpha\rightarrow\nu_\beta)|^2$
is given by
\begin{eqnarray}
\hat{P}(\nu_\alpha\rightarrow\nu_\beta)
&=&|
(N^{*}N^T)_{\alpha\beta}
|^2-4\sum_{j<k}\mbox{\rm Re}(\tilde{X}^{\alpha\beta}_j
\tilde{X}^{\alpha\beta\ast}_k)
\sin^2(\Delta \tilde{E}_{jk}L/2)\nonumber\\
&{\ }&+2\sum_{j<k}\mbox{\rm Im}(\tilde{X}^{\alpha\beta}_j
\tilde{X}^{\alpha\beta\ast}_k)
\sin(\Delta \tilde{E}_{jk}L),
\label{modifiedprob}
\end{eqnarray}
where $\Delta \tilde{E}_{jk}\equiv \tilde{E}_{j}-\tilde{E}_{k}$
and $\tilde{X}^{\alpha\beta}_j\equiv
(N^{*}W)_{\alpha j}(NW^{*})_{\beta j}$~($j=1,2,3)$.
As in the case of the standard neutrino scenario~\cite{Kimura:2002hb,yasuda-kty},
$\tilde{X}^{\alpha\beta}_j$ can be expressed
in terms of the quantity
$X^{\alpha\beta}_j\equiv
U_{\alpha j}^\ast U_{\beta j}$ in vacuum and $\tilde{E}_j$.
To show it, let us first
note the following relations:
\begin{eqnarray}
&{\ }&\sum_j(\tilde{E}_j)^m\tilde{X}^{\alpha\beta}_j
=\sum_j(N^{*}W)_{\alpha j}
(\tilde{E}_j)^m(NW^{*})_{\beta j}
=[
N^{*}({\cal E}+N^T{\cal A}N^{*})^mN^T]_{\alpha\beta}
\equiv Y^{\alpha\beta}_{m+1}\nonumber\\
&{\ }&\qquad\qquad\qquad\qquad\qquad\qquad~~\mbox{\rm for}~m=0,1,2.
\label{simultaneous0}
\end{eqnarray}
Secondly we rewrite Eqs.~(\ref{simultaneous0}) as
\begin{eqnarray}
\sum_{m=1}^3
V_{jm}\,\tilde{X}^{\alpha\beta}_m
=Y^{\alpha\beta}_j\qquad\mbox{\rm for}~j=1,2,3,
\label{simultaneous1}
\end{eqnarray}
where
$V_{jm}\equiv(\tilde{E}_m)^{j-1}$
is the element of 
the van der Monde matrix $V$.
The simultaneous equation~(\ref{simultaneous1}) can be easily solved by inverting $V$:
\begin{eqnarray}
\tilde{X}^{\alpha\beta}_j=\sum_{m=1}^3
(V^{-1})_{jm}\,Y^{\alpha\beta}_m
=\left(\begin{array}{ccc}
(1/\Delta \tilde{E}_{21} \Delta \tilde{E}_{31})
(\tilde{E}_2\tilde{E}_3, & -(\tilde{E}_2+\tilde{E}_3),&
1)\cr
(1/\Delta \tilde{E}_{21} \Delta \tilde{E}_{32})
(\tilde{E}_3\tilde{E}_1, & -(\tilde{E}_3+\tilde{E}_1),&
1)\cr
(1/\Delta \tilde{E}_{31} \Delta \tilde{E}_{32})
(\tilde{E}_1\tilde{E}_2, & -(\tilde{E}_1+\tilde{E}_2),&
1)\cr
\end{array}\right)
\left(\begin{array}{c}
Y^{\alpha\beta}_1\cr
Y^{\alpha\beta}_2\cr
Y^{\alpha\beta}_3
\end{array}\right).\nonumber\\
\label{simultaneous3}
\end{eqnarray}

In the present paper we focus on the short distance
case, i.e., the case where $|\Delta E_{jk}L|\ll 1$,
$|\Delta \tilde{E}_{jk}L|\ll 1$, $|G_Fn_{e,n}L|\ll 1$.
It turns out that terms of order $(\Delta \tilde{E}_{jk}L)^3$ or
higher are negligible at short distance
and we expand the sine functions in the probability~(\ref{modifiedprob}) to quadratic order in the argument
$\Delta \tilde{E}_{jk}L$:
\begin{eqnarray}
\sin^2(\Delta\tilde{E}_{jk}L/2)
\simeq (\Delta\tilde{E}_{jk}L/2)^2,\quad
\sin(\Delta\tilde{E}_{jk}L)
\simeq \Delta\tilde{E}_{jk}L
\nonumber
\end{eqnarray}
Plugging Eq.~(\ref{simultaneous3}) in the probability~(\ref{modifiedprob}) and using the form of $V^{-1}$ in
Eq.~(\ref{simultaneous3}),
we can show that the first terms take the following simple forms:
\begin{eqnarray}
\sum_{j<k}\mbox{\rm Re}(\tilde{X}^{\alpha\beta}_j
\tilde{X}^{\alpha\beta\ast}_k)
(\Delta \tilde{E}_{jk}L/2)^2
&=&(L/2)^2
[\mbox{\rm Re}(Y^{\alpha\beta}_1Y^{\alpha\beta\ast}_3)
-|Y^{\alpha\beta}_2|^2],\nonumber\\
\sum_{j<k}\mbox{\rm Im}(\tilde{X}^{\alpha\beta}_j
\tilde{X}^{\alpha\beta\ast}_k)
\Delta \tilde{E}_{jk}L
&=&-L\,\mbox{\rm Im}(Y^{\alpha\beta}_1Y^{\alpha\beta\ast}_2)
\nonumber
\end{eqnarray}
Hence we obtain the probability at short distance
\begin{eqnarray}
\hat{P}(\nu_\alpha\rightarrow\nu_\beta)
\simeq|(NN^\dagger)_{\alpha\beta}|^2
-L^2[\mbox{\rm Re}(Y^{\alpha\beta}_1Y^{\alpha\beta\ast}_3)
-|Y^{\alpha\beta}_2|^2]
-2L\,\mbox{\rm Im}(Y^{\alpha\beta}_1Y^{\alpha\beta\ast}_2).
\label{modifiedprob2}
\end{eqnarray}

We are interested in the MUV scheme with small values of $\eta$,
where $\eta$ was defined in Eq.~(\ref{N}),
so we evaluate $Y^{\alpha\beta}_j$
only to first order in $\eta$:
\begin{eqnarray}
Y^{\alpha\beta}_1&\simeq&
\delta_{\alpha\beta}+2\eta_{\alpha\beta},\nonumber\\
Y^{\alpha\beta}_2
&\simeq&\sum_{j=2}^3\Delta_{j1}X^{\alpha\beta}_j
+\sum_\gamma\sum_{j=2}^3\Delta_{j1}
(X^{\alpha\gamma}\eta_{\gamma\beta}
+X^{\gamma\beta}\eta_{\alpha\gamma})
+{\cal A}_{\alpha\beta}
+2\eta_{\alpha\beta}({\cal A}_{\alpha\alpha}+{\cal A}_{\beta\beta})
,\nonumber\\
Y^{\alpha\beta}_3|_{\eta\rightarrow0}
&=&\sum_{j=2}^3(\Delta_{j1})^2X^{\alpha\beta}_j
+\sum_{j=2}^3\Delta_{j1}X^{\alpha\beta}_j
({\cal A}_{\alpha\alpha}+{\cal A}_{\beta\beta})
+({\cal A}_{\alpha\beta})^2,
\label{Y}
\end{eqnarray}
where we have used the energy matrix
${\cal E}-E_1{\bf 1}+N^T{\cal A}N^\ast$ instead of~(\ref{energymatrix}) for ease of calculation,
since the shift $(-E_1{\bf 1})$ only changes the phase
of the amplitude $\hat{A}(\nu_\alpha\rightarrow\nu_\beta)$. $Y^{\alpha\beta}_3$ was evaluated only in the limit
$\eta_{\alpha\beta}\rightarrow0$, as it
always appears together with $Y^{\alpha\beta}_1$, which is first
order in $\eta_{\alpha\beta}$.

It is straightforward to obtain
the components $Y^{\mu\tau}_j$ for $j=1,2,3$ from Eqs.~(\ref{Y}).
Keeping only terms to second order in $\sin^2\theta_{13}$, $\Delta_{21}L$, $(\Delta_{31}L)^2$, $(AL)^2$
and first in $\eta_{\alpha\beta}$, and setting $\eta_{e\mu}=0$,
we get the probability
$P_{\mu \tau}=\hat{P}(\nu_\mu\rightarrow\nu_\tau)
/(NN^\dagger)_{\mu\mu}(NN^\dagger)_{\tau\tau}$:
\begin{eqnarray}
P_{\mu \tau} &=& \sin^{2}2\theta_{23}(\Delta_{31} L/2)^2
-2|\eta_{\mu \tau}|\sin\delta_{\mu \tau}\sin2\theta_{23}(\Delta_{31} L)
+4|\eta_{\mu \tau}|^2
\nonumber\\
&-&(1/2)c^2_{12}\sin^22\theta_{23}( \Delta_{31}L)(\Delta_{21} L)-|\eta_{\mu\tau}|\sin2\theta_{23}\cos\delta_{\mu\tau}(AL)(\Delta_{31} L)
\nonumber\\
&+&(1/4)c^4_{12}\sin^22\theta_{23}(\Delta_{21}L)^2+2|\eta_{\mu\tau}|c^2_{12}\sin\delta_{\mu\tau}\sin2\theta_{23}(\Delta_{21} L)
\nonumber\\
&-&(1/4)s_{13}\sin4\theta_{23}\sin2\theta_{12}\cos\delta(\Delta_{31}L)(\Delta_{21} L).
\label{pmutau}
\end{eqnarray}
Notice that matter effects in this channel are clearly subdominant.
All terms but those in the first line of Eq.~(\ref{pmutau}) are numerically smaller than $10^{-4}$ for the setup considered in this work and have thus negligible practical effects.

For the $e\raw\tau$ channel the same formalism has been used to obtain the expanded probability, presented in Eq.~(\ref{probetau}). 

\section*{Acknowledgments}

We acknowledge illuminating discussions with 
Andrea Donini and Pasquale Migliozzi on the efficiencies and backgrounds of the emulsion cloud chamber detector. We are also indebted for stimulating 
comments to Stefan Antusch and Carla Biggio. 
Furthermore, the authors received partial
support from CICYT through the project FPA2006-05423, as well as from
the Comunidad Aut\'onoma de Madrid through Proyecto HEPHACOS;
P-ESP-00346. E.~Fern\'andez-Mart\'{\i}nez acknowledges the UAM for
financial support through a FPU fellowship. J.~L\'opez-Pav\'on acknowledges financial support from the MEC through
the FPU grant with ref. AP2005-1185.
O.~Yasuda was supported in part by Grants-in-Aid for Scientific Research
No.\ 16340078, Japan Ministry
of Education, Culture, Sports, Science, and Technology.


\end{document}